\documentclass[final]{aipproc}

\layoutstyle{6x9}

\usepackage{amsmath,amsfonts,amssymb}

\allowdisplaybreaks

\begin{document}

\title{Extrinsic CPT Violation in Neutrino Oscillations}

\author{Tommy Ohlsson\thanks{In collaboration with: Magnus
  Jacobson. Talk presented at the 5th International Workshop on
  Neutrino Factories \& Superbeams (NuFact'03), Columbia University,
  New York, USA, June 5-11, 2003.}}{
  address={Division of Mathematical Physics, Department of
  Physics, Royal Institute of Technology (KTH) -- Stockholm Center for
  Physics, Astronomy, and Biotechnology (SCFAB), Roslagstullsbacken
  11, SE-106~91~Stockholm, Sweden}
}

\begin{abstract}
In this talk, we investigate extrinsic CPT violation in neutrino
oscillations in matter with three flavors. Note that extrinsic CPT
violation is different from intrinsic CPT violation. Extrinsic CPT
violation is one way of quantifying matter effects, whereas intrinsic
CPT violation would mean that the CPT invariance theorem is not
valid. We present analytical formulas for the extrinsic CPT
probability differences and discuss their implications for
long-baseline experiments and neutrino factory setups.
\end{abstract}

\maketitle

{\it Introduction.}
Recently, there have been several studies on CPT violation in order to
incorporate the results of the LSND experiment \cite{Athanassopoulos:1996jb},
which require a third mass squared difference. However, this is \underline{not}
compatible with three neutrino flavors. Therefore, in most of the
phenomenological studies on CPT violation, different mass squared
differences and mixing parameters for neutrinos and antineutrinos are
introduced by hand leading to four mass squared differences and eight mixing
parameters. Thus, it is possible to include the results of the
LSND experiment.
Note that the results of the LSND experiment will be tested by the MiniBooNE
experiment (September 2002 $\to$ $\sim$ 2005) \cite{miniboone}.
Furthermore, note that another possible description of the results of
the LSND experiment are sterile neutrinos. However, sterile neutrinos
have, in principle, been excluded by the SNO experiment \cite{Bahcall:2002hv}.
Moreover, the first KamLAND data are consistent with the LMA solution
\cite{Bahcall:2002ij}, which means that there is no need for
fundamental CPT violation.

{\it Eccentric or extrinsic CPT violation?}
Let us denote the neutrino oscillation transition probabilities by
$P_{\alpha\beta} \equiv P(\nu_\alpha \to \nu_\beta)$. Then, the CP, T,
and CPT probability differences (pds) are defined as $\Delta
P_{\alpha\beta}^{\rm CP} \equiv P_{\alpha\beta} -
P_{\bar\alpha\bar\beta}$, $\Delta P_{\alpha\beta}^{\rm T} \equiv
P_{\alpha\beta} - P_{\beta\alpha}$, and $\Delta P_{\alpha\beta}^{\rm
CPT} \equiv P_{\alpha\beta} - P_{\bar\beta\bar\alpha}$.

Now, intrinsic (eccentric) CPT violation (or fundamental or genuine
CPT violation) is due to violation of the CPT invariance theorem,
whereas extrinsic CPT violation (or matter-induced or fake CPT
violation) is due to presence of ordinary matter.  Here, we will
assume that the CPT invariance theorem is valid. This implies for the
CP and T pds that the intrinsic and extrinsic effects are mixed,
whereas for the CPT pds there are extrinsic effects only.  Therefore,
non-zero (extrinsic) CPT pds show matter effects, and thus, they are one way of
quantifying such effects.

From conservation of probability, we obtain
$\sum_{\alpha = e,\mu,\tau,\ldots} \Delta P_{\alpha\beta}^{\rm CPT} =
\sum_{\beta = e,\mu,\tau,\ldots} \Delta P_{\alpha\beta}^{\rm CPT}~=~0$.
Note that not all of these equations are linearly independent.
For three neutrino flavors, we have nine CPT pds
for neutrinos. However, only four are linearly independent. Choosing, {\it
e.g.}, $\Delta P_{ee}^{\rm CPT}$, $\Delta P_{e\mu}^{\rm CPT}$,
$\Delta P_{\mu e}^{\rm CPT}$, and $\Delta P_{\mu\mu}^{\rm CPT}$ as the
known ones, the other five can be expressed in terms of these.
Furthermore, we have $\Delta P_{\alpha\beta}^{\rm CPT} = - \Delta
P_{\bar\beta\bar\alpha}^{\rm CPT}$, {\it i.e.}, the CPT pds for
antineutrinos do not give any further information.

{\it The CPT probability differences.}
In vacuum, the CPT pds are
$\Delta P_{\alpha\beta}^{\rm CPT} = 0, \quad \alpha,\beta = e,\mu,\tau$,
whereas, in matter, they are given by
$\Delta P_{\alpha\beta}^{\rm CPT} = | [ S_f(t,t_0) ]_{\beta\alpha} |^2
- | [ {\bar S}_f(t,t_0) ]_{\alpha\beta} |^2$,
where $S_f \equiv S_f(t,t_0)$ and ${\bar S}_f \equiv {\bar
  S}_f(t,t_0)$ are the evolution operators for neutrinos and
antineutrinos, respectively.
We have calculated $S_f$ and ${\bar S}_f$ explicitly using first order
perturbation theory in the small leptonic mixing angle $\theta_{13}$.
These explicit expressions for $S_f$ and ${\bar S}_f$ can be found in
Ref.~\cite{Jacobson:2003wc}.

Two of the CPT pds (with an arbitrary matter density
profile) are:
$\Delta P_{ee}^{\rm CPT} \simeq |\bar\beta|^2 - |\beta|^2$ and
$\Delta P_{e\mu}^{\rm CPT} \simeq c_{23}^2\left(|\beta|^2-|\bar\beta|^2\right)
- 2 c_{23}s_{23} \Im \left(\beta fC - \bar\beta\bar f^\ast \bar A^\ast\right)$,
where $\beta$ and ${\bar \beta}$ describe a part of the two flavor
neutrino evolution in the (1,2)-subsector, $f$ and ${\bar f}$ are some
functions, and ${\bar A}$ and $C$ are complicated
functions that can be found in Ref.~\cite{Jacobson:2003wc}.

In matter of constant density in the low-energy region ($V
\lesssim \delta \ll \Delta$), the CPT pds $\Delta
P_{ee}^{\rm CPT}$ and $\Delta P_{\mu e}^{\rm CPT}$ are calculated to
be \cite{Jacobson:2003wc}
{\small
\begin{eqnarray}
\Delta P_{ee}^{\rm CPT} &\simeq& 8 s_{12}^2 c_{12}^2 \cos 2\theta_{12}
\left( \delta L \cos\frac{\delta L}{2} - 2 \sin\frac{\delta L}{2}
\right) \sin\frac{\delta L}{2} \frac{V}{\delta}
+ \mathcal{O}\left((V/\delta)^3\right), \\
\Delta P_{\mu e}^{\rm CPT} &\simeq& -8 s_{12}^2 c_{12}^2 c_{23}^2 \cos
2\theta_{12} \left( \delta L \cos\frac{\delta L}{2} - 2 \sin\frac{\delta L}{2}
\right) \sin\frac{\delta L}{2} \frac{V}{\delta} \nonumber\\
&-& 16 s_{12} c_{12}^3 s_{13} s_{23} c_{23} \cos
  \delta_{\rm CP} \cos 2\theta_{12} \left( \delta L \cos\frac{\delta
  L}{2} - 2 \sin\frac{\delta L}{2} \right) \sin\frac{\delta L}{2}
  \frac{V}{\delta} \nonumber\\
&+& 16 s_{12} c_{12} s_{13} s_{23} c_{23} \sin
\delta_{\rm CP}
\Bigg\{ \cos 2\theta_{12} \bigg[ \delta L \cos \delta L - \cos \Delta
  L \nonumber\\
&\times& \left( \delta L \cos\frac{\delta L}{2} - 2 \sin\frac{\delta L}{2}
\right) - \sin \delta L \bigg] + \delta L \sin\frac{\delta L}{2} \sin
\Delta L \Bigg\} \frac{V}{\delta} +
\mathcal{O}\left((V/\delta)^3\right),
\end{eqnarray}
}
where $\delta \equiv \frac{\Delta m_{21}^2}{2E_\nu}$, $\Delta \equiv
\frac{\Delta m_{31}^2}{2E_\nu}$, $E_\nu$ is the neutrino energy, $L$
is the baseline length, and $V$ is the matter potential.
Note that if one makes the replacement $\delta_{\rm CP} \to
-\delta_{\rm CP}$, then $\Delta P_{e\mu}^{\rm CPT} \to \Delta P_{\mu
e}^{\rm CPT}$ and $\Delta P_{e\tau}^{\rm CPT} \to \Delta P_{\tau
e}^{\rm CPT}$ and, in the case that $\delta_{\rm CP} = 0$, one has
$\Delta P_{e\mu}^{\rm CPT} = \Delta P_{\mu e}^{\rm CPT}$ and $\Delta
P_{e\tau}^{\rm CPT} = \Delta P_{\tau e}^{\rm CPT}$.
We observe also that in $\Delta P_{ee}^{\rm CPT}$ there are no
$\delta_{\rm CP}$ terms, whereas in $\Delta P_{\mu e}^{\rm CPT}$ there are both
$\sin \delta_{\rm CP}$ and $\cos \delta_{\rm CP}$ terms.
Therefore, it would be possible to extract $\delta_{\rm CP}$ from $\Delta
P_{\mu e}^{\rm CPT}$.
Actually, for symmetric matter density profiles it can be shown that
the $\Delta P_{\alpha\beta}^{\rm CPT}$'s are
always odd functions with respect to the matter potential
$V$ \cite{Minakata:2003,Jacobson:2003wc}.

In the case of a step-function matter density profile in the
low-energy region ($V_{1,2} \lesssim \delta \ll \Delta$), $\Delta
P_{ee}^{\rm CPT}$ is found to be \cite{Jacobson:2003wc}
{\small
\begin{eqnarray}
\Delta P_{ee}^{\rm CPT} &\simeq& 8 s_{12}^2 c_{12}^2 \cos 2\theta_{12}
\bigg[ \delta \left( L_1 \frac{V_1}{\delta} + L_2
\frac{V_2}{\delta} \right) \cos \frac{\delta (L_1 + L_2)}{2} \nonumber\\
&-& 2 \left( \frac{V_1}{\delta} \sin \frac{\delta L_1}{2} \cos
  \frac{\delta L_2}{2} + \frac{V_2}{\delta} \sin \frac{\delta L_2}{2}
  \cos \frac{\delta L_1}{2} \right) \bigg] \sin \frac{\delta (L_1 +
  L_2)}{2}.
\end{eqnarray}
}
Note that this formula is completely symmetric with respect to
interchange of layers 1 and 2 and, in the limit $V_{1,2} \to V$ and
$L_{1,2} \to L/2$, one has $\Delta P_{ee}^{\rm
CPT}(\mbox{step-function}) \to \Delta P_{ee}^{\rm CPT}(\mbox{constant})$.

Similarly, in the case of the T probability difference in the
low-energy region ($\delta = \Delta m_{21}^2/(2E_\nu) \gtrsim
V_{1,2}$), one finds \cite{Akhmedov:2001kd}
{\small
\begin{eqnarray}
\Delta P_{\alpha\beta}^{\rm T} &\simeq& \cos \delta_{\rm CP} \cdot 8
\underbrace{s_{12} c_{12} s_{13} s_{23} c_{23}
\frac{\sin(2\theta_1 - 2\theta_2)}{\sin 2\theta_{12}}}_{J_{\rm eff}}
\left\{ s_1 s_2 \left[Y - \cos \left( \Delta_1 L_1 + \Delta_2
L_2 \right) \right] \right\} \nonumber\\
&+& \sin \delta_{\rm CP} \cdot 4 s_{13} s_{23}
c_{23} X_1 \left[ Y - \cos
\left( \Delta_1 L_1 + \Delta_2 L_2 \right) \right],
\end{eqnarray}
}
where $J_{\rm eff}$ is an effective Jarlskog invariant. (See
Ref.~\cite{Akhmedov:2001kd} for the definitions of the different quantities in
this formula.)
Here the $\cos \delta_{CP}$ term is due to matter-induced T violation,
whereas the usual $\sin \delta_{CP}$ term is due to fundamental T violation.

\begin{figure}[t!]
\includegraphics[height=8.65cm,angle=-90]{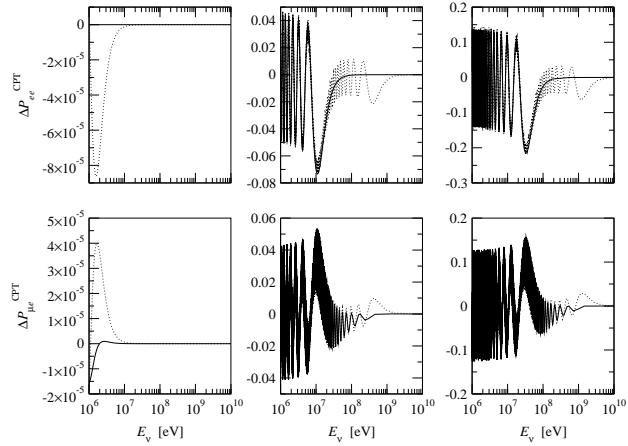}
\caption{\label{fig:CPTprobs} The CPT pds $\Delta
  P_{ee}^{\rm CPT}$ and $\Delta P_{\mu e}^{\rm CPT}$ as functions of
  the neutrino energy $E_\nu$ for baseline lengths $L \in \{1,250,750\} \, {\rm
  km}$. The solid curves show analytical results, whereas the dotted
  curves show numerical results. The fast oscillations present in the
  numerical results are averaged out in the analytical calculations.}
\end{figure}

{\it Numerical calculations and implications.}
Using the present best-fit values of the fundamental neutrino parameters,
$\Delta m_{21}^2 \simeq 7.1 \cdot 10^{-5} \, {\rm eV}^2$, $|\Delta
m_{31}^2| \simeq 2.5 \cdot 10^{-3} \, {\rm eV}^2$, $\theta_{12} \simeq
34^\circ$, $\theta_{23} \simeq 45^\circ$, and, in addition, choosing a
normal mass hierarchy spectrum [${\rm sgn}(\Delta m_{31}^2) = 1$],
$\theta_{13} = 9.2^\circ$, and $\delta_{\rm CP} = 0$, we
have calculated the CPT pds for some past, present, and future long-baseline
experiments. The results are presented in Table~\ref{tab:experiments}.
\begin{table}[t!]
\begin{tabular}{lcc}
\hline
Experiment & \multicolumn{2}{c}{CPT probability differences}\\
 & Quantities & Numerical value\\
\hline
BNL NWG & $\Delta P_{\mu e}^{\rm CPT}$ & $0.010$\\
BNL NWG & $\Delta P_{\mu e}^{\rm CPT}$ & $0.032$\\
BooNE & $\Delta P_{\mu e}^{\rm CPT}$ & $6.6 \cdot 10^{-13}$\\
MiniBooNE & $\Delta P_{\mu e}^{\rm CPT}$ & $4.1 \cdot 10^{-14}$\\
CHOOZ & $\Delta P_{ee}^{\rm CPT}$ & $-3.6 \cdot 10^{-5}$\\
ICARUS & $\Delta P_{\mu e}^{\rm CPT}$ & $4.0 \cdot 10^{-5}$\\
 & $\Delta P_{\mu\tau}^{\rm CPT}$ & $-3.8 \cdot 10^{-5}$\\
JHF-Kamioka & $\Delta P_{\mu e}^{\rm CPT}$ & $3.8 \cdot 10^{-3}$\\
 & $\Delta P_{\mu\mu}^{\rm CPT}$ & $-1.3 \cdot 10^{-4}$\\
K2K & $\Delta P_{\mu e}^{\rm CPT}$ & $1.0 \cdot 10^{-3}$\\
 & $\Delta P_{\mu\mu}^{\rm CPT}$ & $-5.3 \cdot 10^{-5}$\\
\hline
\end{tabular}
\begin{tabular}{lcc}
\hline
Experiment & \multicolumn{2}{c}{CPT probability differences}\\
 & Quantities & Numerical value\\
\hline
KamLAND & $\Delta P_{ee}^{\rm CPT}$ & $-0.033$\\
LSND & $\Delta P_{\mu e}^{\rm CPT}$ & $4.8 \cdot 10^{-15}$\\
MINOS & $\Delta P_{\mu e}^{\rm CPT}$ & $1.9 \cdot 10^{-4}$\\
 & $\Delta P_{\mu\mu}^{\rm CPT}$ & $-1.1 \cdot 10^{-5}$\\
NuMI I & $\Delta P_{\mu e}^{\rm CPT}$ & $0.026$\\
NuMI II & $\Delta P_{\mu e}^{\rm CPT}$ & $2.6 \cdot 10^{-3}$\\
NuTeV & $\Delta P_{\mu e}^{\rm CPT}$ & $1.6 \cdot 10^{-18}$\\
NuTeV & $\Delta P_{\mu e}^{\rm CPT}$ & $8.2 \cdot 10^{-20}$\\
OPERA & $\Delta P_{\mu\tau}^{\rm CPT}$ & $-3.8 \cdot 10^{-5}$\\
Palo Verde & $\Delta P_{ee}^{\rm CPT}$ & $-1.2 \cdot 10^{-5}$\\
Palo Verde & $\Delta P_{ee}^{\rm CPT}$ & $-2.2 \cdot 10^{-5}$\\
\hline
\end{tabular}
\caption{\label{tab:experiments} Extrinsic CPT pds for some
  past, present, and fututre long-baseline experiments.}
\end{table}
Note that $\Delta P_{ee}^{\rm CPT}$ for the
KamLAND experiment is $|\Delta P_{ee}^{\rm CPT}| \sim 3~\%$,
which means that extrinsic CPT violation is non-negligible for
this experiment. The problem is just how one should obtain $P_{ee}$
for the same neutrino energy and baseline length.
Furthermore, we have calculated $\Delta
P_{\mu e}^{\rm CPT}$ for two neutrino factory setups, using
the following parameter values: $\rho = \rho_{\rm mantle}
\simeq 4.5 \, {\rm g/cm^3}$, $E_\nu = 50 \, {\rm GeV}$, $L \in
\{3~000, 7~000\} \, {\rm km}$. Then, $L = 3~000 \, {\rm
km}$ leads to $\Delta P_{\mu e}^{\rm CPT} \simeq 3.0 \cdot 10^{-5}$,
whereas $L = 7~000 \, {\rm km}$ leads to $\Delta P_{\mu
e}^{\rm CPT} \simeq 1.8 \cdot 10^{-5}$. Therefore, extrinsic CPT
violation is practically negligible for a future neutrino factory.
In Fig.~\ref{fig:CPTprobs}, $\Delta
P_{ee}^{\rm CPT}$ and $\Delta P_{\mu e}^{\rm CPT}$ are shown plotted
as functions of $E_\nu$ for $L \in \{ 1,250,750 \}
\, {\rm km}$.
We note that an increasing $L$ implies an increasing
values of the $\Delta P_{\alpha\beta}^{\rm CPT}$'s and when 
$E_\nu \to \infty$ we observe that $\Delta P_{\alpha\beta}^{\rm CPT} \to 0$.

{\it Summary \& Conclusions.}
In conclusion, we have studied extrinsic CPT violation in three flavor neutrino
oscillations assuming the CPT invariance theorem. In general, the
(extrinsic) CPT pds for an arbitrary matter density profile have been
derived. In particular, first order perturbation theory formulas for
constant and step-function matter density profiles have been
calculated as well as low-energy approximations. Furthermore,
implications for accelerator and reactor long-baseline experiments as
well as neutrino factory setups have been presented. For certain
experiments the CPT pds can be as large as $|\Delta
P_{\alpha\beta}^{\rm CPT}| \sim 5~\%$. In general, the CPT pds increase with
increasing baseline length and decrease with increasing neutrino energy.

{\it Acknowledgments.}
I would like to thank S.M.~Bilenky, M.~Jacobson, R.~Johansson, 
M.~Lindner, H.~Minakata, G.~Seidl, H.~Snellman, and W.~Winter for
useful discussions and comments. 
This work was supported by the Swedish Research Council
(Vetenskapsr{\aa}det), Contract No.~621-2001-1611, 621-2002-3577, the
Magnus Bergvall Foundation (Magn.~Bergvalls Stiftelse), and the
Wenner-Gren Foundations.

\IfFileExists{\jobname.bbl}{}
 {\typeout{}
  \typeout{******************************************}
  \typeout{** Please run "bibtex \jobname" to optain}
  \typeout{** the bibliography and then re-run LaTeX}
  \typeout{** twice to fix the references!}
  \typeout{******************************************}
  \typeout{}
 }

\end{document}